\documentclass[10pt,conference]{IEEEtran}
\IEEEoverridecommandlockouts
\usepackage{algorithmic}
\usepackage{graphicx}
\usepackage{textcomp}
\usepackage{xcolor}
\usepackage{booktabs} 
\usepackage{mdframed}
\usepackage{xspace}
\usepackage[normalem]{ulem}
\useunder{\uline}{\ul}{}
\usepackage{multirow}
\usepackage{lscape}
\usepackage{rotating}
\usepackage{supertabular}
\usepackage{longtable}
\usepackage{enumerate}
\usepackage{colortbl}
\usepackage{subcaption}
\usepackage{hhline}
\usepackage{makecell}
\usepackage{caption}
\usepackage{hyperref}
\usepackage{enumitem}
\usepackage{tcolorbox}
\usepackage{tabularx}
\usepackage{lipsum}
\usepackage{balance}
\usepackage{blindtext,graphicx}
\usepackage[absolute]{textpos}
\usepackage{fdsymbol}
\usepackage{url}

\usepackage{breakurl}
\usepackage{float}
\usepackage{pgfplots}
\usepackage{pgf-pie}  

\usepackage{fontawesome}
\AtBeginDocument{%
  \providecommand\BibTeX{{%
    \normalfont B\kern-0.5em{\scshape i\kern-0.25em b}\kern-0.8em\TeX}}}

\begin{document}

\title{How LLMs Aid in UML Modeling: An Exploratory Study with Novice Analysts}

\makeatletter 
\newcommand{\linebreakand}{%
  \end{@IEEEauthorhalign}
  \hfill\mbox{}\par
  \mbox{}\hfill\begin{@IEEEauthorhalign}
}
\makeatother 

\author{\IEEEauthorblockN{Beian Wang}
\IEEEauthorblockA{\textit{School of Computer Science} \\
\textit{Wuhan University}\\
Wuhan, China \\
wba20021119@outlook.com}
\and
\IEEEauthorblockN{Chong Wang*}
\IEEEauthorblockA{\textit{School of Computer Science} \\
\textit{Wuhan University}\\
Wuhan, China \\
cwang@whu.edu.cn}
\and
\IEEEauthorblockN{Peng Liang}
\IEEEauthorblockA{\textit{School of Computer Science} \\
\textit{Wuhan University}\\
Wuhan, China \\
liangp@whu.edu.cn}
\and
\linebreakand
\IEEEauthorblockN{Bing Li}
\IEEEauthorblockA{\textit{School of Computer Science} \\
\textit{Wuhan University}\\
Wuhan, China \\
bingli@whu.edu.cn}
\and
\IEEEauthorblockN{Cheng Zeng}
\IEEEauthorblockA{\textit{School of Computer Science} \\
\textit{Wuhan University}\\
Wuhan, China \\
zengc@whu.edu.cn}
}

\maketitle

\begin{abstract}
Since the emergence of GPT-3, Large Language Models (LLMs) have caught the eyes of researchers, practitioners, and educators in the field of software engineering. However, there has been relatively little investigation regarding the performance of LLMs in assisting with requirements analysis and UML modeling. This paper explores how LLMs can assist novice analysts in creating three types of typical UML models: use case models, class diagrams, and sequence diagrams. For this purpose, we designed the modeling tasks of these three UML models for 45 undergraduate students who participated in a requirements modeling course, with the help of LLMs. By analyzing their project reports, we found that LLMs can assist undergraduate students as novice analysts in UML modeling tasks, but LLMs also have shortcomings and limitations that should be considered when using them.
\end{abstract}



\maketitle

\section{Introduction}


With the emergence of GPT-3, the Large Language Models (LLMs for short) have made remarkable improvements in both performance and versatility. This brings breakthroughs to natural language processing (NLP), computer vision, and various other domains. Especially, in Software Engineering (SE), LLMs have been effectively employed in code generation~\cite{liu2024your}~\cite{vaithilingam2022expectation}, software architecture~\cite{ahmad2023towards}, software testing~\cite{zimmermann2023automating}, etc. 
As an important and early stage in the software development life cycle, requirements engineering (RE) aims to discover, understand, formulate, analyze, and agree on what problem should be solved, why such a problem needs to be solved, and who should be involved in the responsibility of solving problems~\cite{lamsweerde2009requirements}. Generally, software requirements are described with natural languages or graphic models, and requirements modeling is a human-intensive and time-consuming task. The outstanding capability of LLMs in NLP and their interaction mode make them highly promising for writing requirements in natural languages. However, a few studies investigated the capability of LLMs in generating graphical requirements models. In SE, UML (Unified Modeling Language) has been widely used to help in requirements analysis and modeling with a set of graphic notations for researchers, educators, and practitioners. Therefore, it is meaningful to explore how LLMs aid in the UML modeling task for RE and to evaluate the capability of LLMs in requirements modeling tasks.  


For this purpose, this paper explored the performance of LLMs aiding in creating three types of UML models, i.e., use case diagrams, class diagrams, and sequence diagrams. More specifically, we designed and conducted an experiment in our course, i.e., software requirements analysis and modeling. 45 undergraduate students majoring in Software Engineering at Wuhan University joined in this course were asked to create UML use case diagrams, class diagrams, and sequence diagrams for a given case study~\cite{replpack}, with the help of LLMs. By manually analyzing the created UML models as well as the human-LLM conversation for this modeling tasks in their project reports, we found that LLM can aid in the creation and optimization of these three types of UML models, but LLMs still have some shortcomings and limitations. These findings may provide insights for practitioners in requirements engineering and software education professionals in the field of requirements analysis and UML modeling. 

The main contributions of this paper are as follows. 
\begin{itemize}
    \item The effectiveness of LLMs in creating three types of UML models, i.e. use case diagrams, class diagrams, and sequence diagrams, was investigated and evaluated with the course projects in the undergraduate program in university. 
    \item The evaluation criteria for the created three types of UML models with LLMs were proposed in detail to cover the core modeling elements of these three UML models. 
    \item The output formats of the LLM-aided UML models were summarized to explore the impact factors of applying LLMs in the UML modeling tasks. 
\end{itemize}

The remainder of this paper is organized as follows. Section~\ref{sec:relatedwork} introduces the background and relevant studies on this topic. Section~\ref{sec:rprocess} defines the research questions and presents the details of our experimental design. Section~\ref{sec:results} provides the experimental results, and Section~\ref{sec:discussion} discusses these results. Section~\ref{sec:limitations} addresses the limitations of this study. Section~\ref{sec:conclusion} concludes the work and future research directions.

\section{Related Work}\label{sec:relatedwork}
With the development of GPT-3, LLMs have sparked widespread interest in various academic and industrial domains. The use of LLMs in SE tasks is now becoming a hot research topic~\cite{zheng2023understanding}. 

In the SE community, researchers have conducted some studies using large language models, primarily focusing on code generation and software architecture assistance. For example, Ahmad et al. designed a case study that involves collaboration between a novice software architect and ChatGPT to architect a service-based software system. The case study reflects the factors that need to be considered in collaborative architecture design with ChatGPT, such as Variation in responses and artifacts, level of human decision support~\cite{ahmad2023towards}. White et al. provided a catalog of patterns for software engineering that classifies patterns according to the types of problems they solve. They also explored several prompt patterns that had been applied to improve requirements elicitation, rapid prototyping, code quality, refactoring, and system design~\cite{white2023chatgpt}. Xie et al. conducted the first empirical study to evaluate the capabilities of LLMs to generate software specifications from software comments or documentation. The results showed that with few shot learning, LLMs outperformed the  traditional methods and more sophisticated prompt construction strategies can further increase this performance gap~\cite{xie2023impact}. Jeuring et al. explored whether computational thinking (CT) skills can predict the ability to develop software using LLM-based tools in their experiment, and the results showed that the ability to develop software using LLM-based tools can indeed be predicted by the score on a CT assessment ~\cite{jeuring2023skills}. Sadik et al. provided an Agile Model-Driven Development (MDD) approach to enhance code auto-generation using GPT-4, and the results show that the auto-generated code was aligned with the expected UML sequence diagram~\cite{sadik2023coding}.

As for RE, there is also an amount of researches about the application of LLMs across various stages of the RE process. Luitel et al. explored whether LLMs are useful external sources of knowledge for the detection of potential incompleteness in natural language requirements by utilizing BERT. They found that BERT-based predictions effectively highlight terminology that is missing from requirements descriptions~\cite{luitel2023improving}. Zhang et al. empirically evaluated the performance of ChatGPT in requirements information retrieval (IR) tasks to derive insights into the design or development of more effective requirements retrieval methods or tools based on generative LLMs~\cite{zhang4450322evaluation}. Waseem et al. investigated how novice developers use ChatGPT during the software development life cycle (SDLC), including software design using ChatGPT and PlantUML. The results revealed that ChatGPT has a positive influence on various development phases of SDLC, with enhanced efficiency, accuracy, and collaboration in software development~\cite{waseem2023using}. Arora et al. explored the use of LLMs in driving RE processes, focusing on the potential for requirements elicitation, analysis, specification, and validation, and found that LLMs can enhance several RE tasks by automating, streamlining, and augmenting human capabilities~\cite{arora2023advancing}. Kanuka et al. explored the feasibility of using ChatGPT to create bidirectional traceability between design models (e.g., UML class diagrams) and Code, and the study results show that ChatGPT is capable of generating design models and code from natural language requirements, and link them~\cite{kanuka2023exploring}.

However, only a few researcher employed LLMs in requirements analysis and modeling, especially in UML modeling. For instance, Cámara et al. investigated the strengthens and weakness of ChatGPT in helping with modeling tasks~\cite{camara2023assessment}. However, the authors only concentrated on UML class diagrams enriched with OCL constraints. Our work covers three types of UML models that are frequently used in requirements modeling. Furthermore, we attempt to identify the factors that influence LLM performance in the requirements modeling tasks with UML, such as output formats.

\section{Research Process} \label{sec:rprocess}

\begin{figure*} [t]
    \centering   
    \includegraphics[width=1\linewidth]{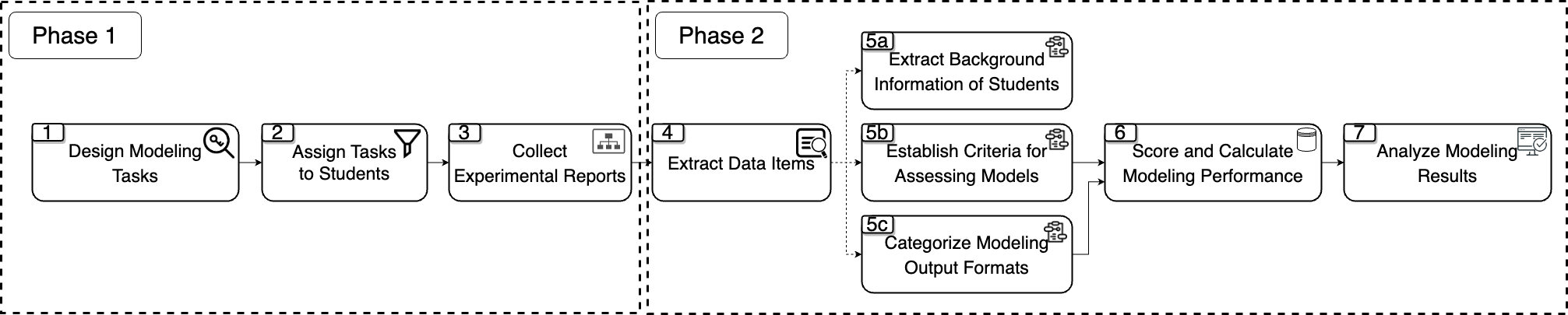}
    \vspace*{-0.9em}
    \caption{The process of the experiment}   
    \label{fig:draw}
    \vspace*{-1.6em}
\end{figure*}

\subsection{Research Objective}
The main objective of this paper is to explore the performance of LLMs in requirements analysis and modeling with UML. Generally, UML use case models, class diagrams, and sequence diagrams are the three essential and most widely used UML models in object-oriented requirements analysis and design~\cite{Larman2012Applying}. Therefore, this study mainly focuses on the performance of LLMs in aiding in the creation of these three types of UML models.

\subsection{Research Questions}
To achieve the research objective, we formulate the following four research questions (RQs). 

\textbf{RQ1}: \textit{How do LLMs perform in identifying modeling elements in UML use case diagram?}

\textbf{Rationale:} This RQ is designed to explore whether LLMs can aid in creating UML use case models as well as their strengths and/or weaknesses in this modeling task. The answer of this RQ would help us understand the capability of LLMs to identify basic modeling elements in the use case models.

\textbf{RQ2}: \textit{How do LLMs perform in identifying modeling elements in UML class diagram?}

\textbf{Rationale:} This RQ is designed to explore whether LLMs can aid in creating UML class diagrams as well as their strengths and/or weaknesses in this modeling task. Answering this RQ would help understand the capability of LLMs to identify basic modeling elements in the UML class diagrams.

\textbf{RQ3}: \textit{How do LLMs perform in identifying modeling elements in the UML sequence diagram?}

\textbf{Rationale:} This RQ is designed to explore whether LLMs can aid in creating UML sequence diagrams as well as their strengths and/or weaknesses in this task. The answer of this RQ would help us understand the capability of LLMs to identify basic modeling elements in the UML sequence diagrams.

\textbf{RQ4}: \textit{Does the specified output format affect the performance of LLMs in UML modeling?}

\textbf{Rationale:} In the project reports, some participants asked the LLM to return the specified formats, such as UML models or the PlantUML code, for the given case study. This RQ intends to explore what formats can be returned by the LLM and whether different formats of these LLM outputs affect the LLM's performance in UML modeling. The answer to this RQ would help us understand the types of LLM output formats and which LLM output formats fit the expected UML models most.  




\subsection{Experiment Design}
To answer these four RQs, this sub-subsection designed a series of experiments, consisting of the following two phases. The entire process of the experiment can be referenced in Figure~\ref{fig:draw}.

\subsubsection{Phase 1: Setting up modeling tasks}
This phase aims to guide and conduct a specified UML modeling task aided by LLMs, referring to Step 1 to Step 3 in Figure~\ref{fig:draw}. More specifically, the participants are required to create three types of UML models, i.e., a use case model, a class diagram, and a sequence diagram, for the given case study: Order Processing System (see the case study description in~\cite{replpack}). Furthermore, the LLMs, such as ChatGPT, should be employed to complete this UML modeling task, in order to examine the performance of LLM in assisting the creation of these three types of models.

The participants in this study are 45 undergraduate students majoring in Software Engineering (SE) from the School of Computer Science of our university. From February to May, 2023, they have not only learned knowledge but also got training on how to create the UML use case models, class diagrams, and sequence diagrams for a given case study during their requirements analysis and modeling course. All the participants are required to deliver their project reports in the middle of June, 2023. The report consists of two parts: (1) the UML use case model, class diagram, and sequence diagram created for the given case study in~\cite{replpack}, with the help of the selected LLM; (2) the history of the interaction between the student and the LLM.

\begin{table*}[h]
\caption{Data Items Extracted and Their related RQs}
\label{tab:dataitems}
\begin{tabular}{lp{4.5cm}p{9cm}p{2cm}}
\hline
\textbf{\#} & \textbf{Data Item}     & \textbf{Description}       & \textbf{Related RQ(s)} \\ \hline
\hline
D1 & Background Information & Some basic information about participants  & Demographics  \\ \hline
D2  &  Modeling elements in use case models   & Actors, use cases, the relationships between actors, the relationship between actor and use cases, etc., in the UML use case diagrams. & RQ1\\ \hline
D3  &  Modeling elements in class diagrams & Class as well as its attributes and operations, the relationships between classes, etc., in the UML class diagrams. & RQ2\\ \hline
D4   &  Modeling elements in sequence diagrams  & Objects, messages sent from one object to another, the order of messages, etc., in the UML sequence diagram. & RQ3\\ \hline
D5 & Formats of UML models & The notations supported by the specified tools and used to describe the UML models in the project reports. & RQ4 \\ \hline
\end{tabular}
\end{table*}

\subsubsection{Phase 2: Evaluate modeling results}
In this phase, we intend to analyze the collected 45 project reports, including the UML models created with the help of LLMs and the details of the human-LLM conversation. Phase 2 covers Step 4 to Step 7 in Figure ~\ref{fig:draw}. To evaluate the modeling results, we extracted some data items from each report, as shown in Table ~\ref{tab:dataitems}.

\textbf{D1} is collected to understand the background and experience on using LLMs of the participants, including the following information in each report: \begin{itemize}
  \item Participant ID
  \item Participant's prior experience with LLMs
  \item The LLM used in this UML modeling task
  \item Language of prompts
\end{itemize}

\textbf{D2, D3, and D4} refer to the three types of UML models created by participants with the assistance of LLM. To evaluate the quality of these models, we defined a set of criteria to calculate the scores of the created models. Table~\ref{tab:uccriteria}, ~\ref{tab:ccriteria} and ~\ref{tab:scriteria} list the evaluation criteria for the created use case model, class diagram, and sequence diagram respectively. 

\begin{table}[h]
\small
\caption{Evaluation criterion for the created use case model}
\label{tab:uccriteria}
\centering
\begin{tabular}{lp{2cm}p{5cm}}
\hline
No. & \textbf{Criteria} & \textbf{Scoring} \\ \hline
\hline
UC1 & Complete identification of actors &
  Primary actors should align with the textual description, be essential in existence, and have appropriate names. \newline
  \textbf{To obtain the score}, primary actors must be included. Different expressions with similar meanings are allowed. \newline
\textbf{Primary actors}: Customer, Gold Customer, Accounting System
 \\ \hline
UC2 & Complete identification of use cases &
  The created use cases are reflected in the text and can fully describe the system processes. \newline 
  \textbf{To obtain the score}, core use cases must be included. Different expressions with similar meanings are allowed. \newline
\textbf{Core use cases}: Place Order, Check Order Status, Cancel Order (or Modify Order), Signup for Notification, Return Product
  \\ \hline
UC3 & Accurate recognition of relationships among actors &
  The relationship among actors are correctly identified. \newline
  \textbf{To obtain the score}, the generalization relationship between `Gold Customer' and `Customer' should be accurately represented. \\ \hline
UC4 & Accurate recognition of relationships between actors and use cases &
  Actors and use cases are correctly connected, aligning with the description provided in the prompt. \newline
  \textbf{To obtain the score}, use cases must be connected to the corresponding actors, with the correct UML notations. 
  \\ \hline
\end{tabular}%
\end{table}

\begin{table}[h]
\small
\caption{Evaluation criterion for the created UML class diagram}
\label{tab:ccriteria}
\centering
\begin{tabular}{lp{2cm}p{5cm}}
\hline
No. & \textbf{Criteria} & \textbf{Scoring} \\ \hline
\hline
CC1 & Correct identification of classes &
 All necessary concepts and relationships mentioned in the description are adequately covered. There are no errors or logical inconsistencies. The names of classes are clear and descriptive enough. \newline
 \textbf{To obtain the score}, core classes must be included. Different expressions with similar meanings are allowed. \newline
\textbf{Core classes}: Customer, Gold Customer, Order, OrderProcessingManager, Product, Invoice, ShippingInfo
 \\ \hline
CC2 & Correct identification of operations &
  Operations are contained in the corresponding classes. The included operations should be named appropriately to avoid conflicts with the class and generally adhere to object-oriented design principles. \newline
  \textbf{To obtain the score}, the class diagram should well-structured and reasonable. It will be manually reviewed and evaluated by the authors of this paper. 
  \\ \hline
CC3 & Correct identification of attributes &
  Attributes are contained within the classes. The names of the attributes do not conflict with the class. 
  \newline
  \textbf{To obtain the score}, the attributes should be defined with appropriate names and fit the class they belong to. It will be manually reviewed and evaluated by the authors of this paper.
  \\ \hline
CC4 & Accurate recognition of relationships among classes &
  The relationships between classes are correctly and completely represented, such as inheritance, association, aggregation, composition, etc. There are no evident errors in the relationships. \newline
  \textbf{To obtain the score}, the core relationships must be accurately represented. Besides, there should be no obvious commonsense or logical errors in the relationships between classes. \newline
  \textbf{Core relationships:} The relationship between Customer and Golden Customer, OrderProcessingManager and Order, GoldCustomer and BackOrderNotifier.
  \\ \hline
\end{tabular}%
\end{table}

\begin{table}[h]
\small
\caption{Evaluation criteria for the created UML sequence diagram}
\label{tab:scriteria}
\centering
\begin{tabular}{lp{2cm}p{5cm}}
\hline
No. & \textbf{Criterion} & \textbf{Description} \\ \hline
\hline
SC1 & Correct identification of objects &
  All necessary interactions and objects are included, and there are no redundant or unnecessary objects. The extraction and creation of objects align with the textual description, with no evident logical errors. \newline
  \textbf{To obtain the score,} core objects must be included. Different expressions with similar meanings or functions are allowed. \newline
  \textbf{Core objects}: (All) Customer, PlaceOrderScreen+OrderProcessingManager (or OrderSystem), InventoryManager
  \\ \hline
SC2 & Correct identification of messages &
  Message names are clear, descriptive, and accurately representing the message type and purpose. Message definitions align with the textual description, with no evident logical or common-sense violations. \newline
  \textbf{To obtain the score}, the source and target objects of the message as well as the names of the messages should be reasonable. The identification of messages will be manually reviewed and evaluated by the authors of this paper.
  \\ \hline
SC3 & Correct identification of the order of messages &
  The message sequence is reasonable, providing a complete description of the system process as presented in the text, with no evident logical or common-sense errors. \newline
  \textbf{To obtain the score}, the order of the messages should be reasonable. The order of messages will be manually reviewed and evaluated by the authors of this paper.
  \\ \hline
\end{tabular}%
\end{table}

To simplify the evaluation process, the score of each of these eleven criteria is set as either 0 or 1. If the modeling elements follow certain criteria in these three tables, it earns one point; otherwise, 0 point. The specific scoring method (such as whether core elements are covered, meeting certain criteria, or reviewed by UML modeling experts (i.e., the authors of this paper)) are also presented in Table~\ref{tab:uccriteria}, ~\ref{tab:ccriteria}, and ~\ref{tab:scriteria}. In this way, the ranges of scores for UML use case model, class diagram, and sequence diagram are from 0 to 4,0 to 4, and 0 to 3 respectively. 

Formula~(\ref{eq:eq1}) is defined to calculate the correctness rate of each criterion in these three tables, i.e., $CR(UC_i)$,~$CR(CC_j)$, and~$CR(SC_k)$, after grading the UML models in all the 45 project reports.

\begin{equation}
\label{eq:eq1}   
\begin{aligned}
   & CR(UC_i|CC_j|SC_k) = \\
   & \frac{\sum_{m=1}^{N}Score(UC_i)_m|Score(CC_j)_m|Score(SC_k)_m}{N} 
\end{aligned}
\end{equation}

where $Score(UC_i)_m$,~$Score(CC_j)_m$, and~$Score(SC_k)_m$ denotes the score of the $i$th use case model criterion, the~$j$th class diagram criterion, and the~$k$th sequence model criterion when evaluating the corresponding UML models in the $m$th participant's project report. In this work, $i\in[1,4]$,~$j\in[1,4]$,~$k\in[1,3]$, and~$N=45$.




For each type of the three UML model, Formula~(\ref{eq:eq2}) is defined to respectively calculate its average score, i.e., $Avg(UM)$,~$Avg(CM)$, and~$Aug(SM)$, and to evaluate the LLM's performance in creating this type of model.

\begin{equation}
\label{eq:eq2}
Avg(UM|CM|SM) = \frac{\sum_{i|j|k=1}^{n_u|n_c|n_s}CR(UC_i|CC_j|SC_k)}{n_u|n_c|n_s}  
\end{equation}

$n_(u|c|s)$ represents the number of criteria for each UML model. In this experiment, $n_u=4$, ~$n_c=4$, and ~$n_s=3$.

Both \textbf{D5} and \textbf{D6} are extracted from the human-LLM conversation record documented in the 45 project reports. Different conversations bring different UML models in various formats for the same case study. To evaluate the impact of different formats (i.e., \textbf{D5}) on the LLM's performance, we first define Formula~(\ref{eq:eq3}) to calculate the total score of the three UML models (that is, $Sum_m$) in each project report. According to the formats of the included UML models, the 45 project reports are grouped into several categories to get their average scores respectively. A higher average score means a better average quality of a specific type of output formats.

\begin{equation}
\label{eq:eq3}
\begin{aligned}
& Sum_m=\sum_{i=1}^{n_u}Score(UC_i)_m+\sum_{j=1}^{n_c}Score(CC_j)_m \\
&       +\sum_{k=1}^{n_k}Score(SC_k)_m   
\end{aligned}
\end{equation}

Similarly, the interaction patterns (i.e., \textbf{D6}) adopted in the human-LLM conversations are identified and categorized, to explore the impact of different ways of interacting with the LLM on the quality of the three generated UML models.





\begin{figure}[h]
    \includegraphics[width= 0.9\columnwidth]{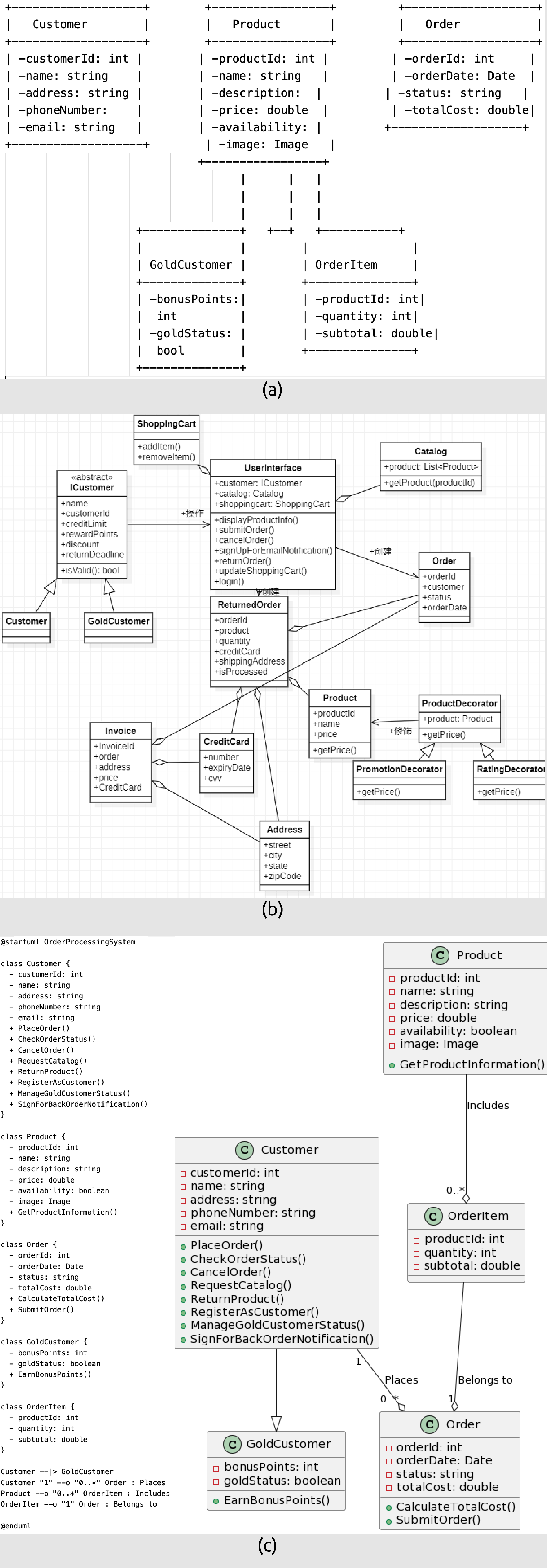}
    \caption{Examples of the three output formats of LLMs}
    \label{fig:abc}
\end{figure}

\section{Results}\label{sec:results}

\subsection{Demographics of the Participants} \label{sec:DoP}


As shown in Figure~\ref{fig:pie1}, 66.7\% (30 out of 45) participants have prior knowledge and/or experience in utilizing LLMs. In particular, 56.7\% (17 out of these 30) participants have experience on using specific LLM for requirements analysis and modeling in their previous projects and/or courses.

\begin{figure} [h]
    \centering   
    \includegraphics[width=0.95\linewidth]{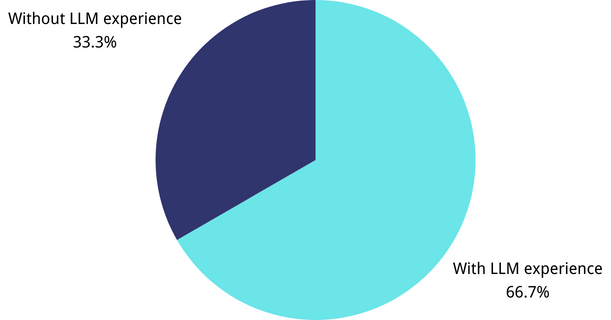}
    \caption{Distribution of the participants with/without experience of using LLMs}   
    \label{fig:pie1}
\end{figure}

Regarding the LLMs applied in our UML modeling task, ChatGPT (including ChatGPT 3.5 and ChatGPT 4.0) is the most used one, accounting for 93.3\% of the 45 participants (42 participants), as presented in Figure ~\ref{fig:pie2}. The remaining three participants employed NewBing in their UML modeling tasks. 

\begin{figure} [h]
    \centering   
    \includegraphics[width=0.5\linewidth]{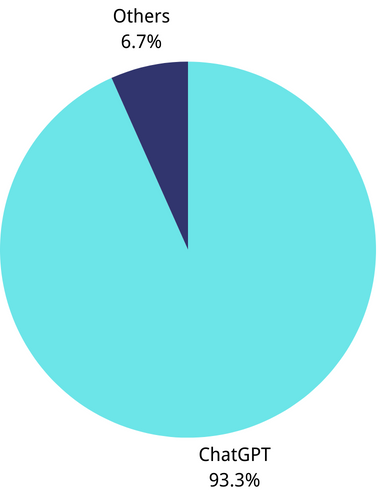}
    \caption{Distribution of the LLMs used in UML modeling tasks}   
    \label{fig:pie2}
\end{figure}

During the human-LLM interaction, 68.9\% of the 45 participants (31 participants) interacted with the LLM in English, and the remaining 14 participants used Chinese, as shown in Figure ~\ref{fig:pie3}. 

\begin{figure} [h]
    \centering   
    \includegraphics[width=0.65\linewidth]{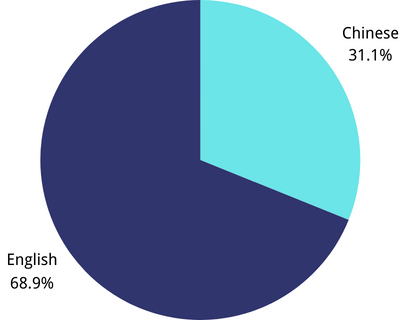}
    \caption{Distribution of the languages used in the human-LLM interaction}   
    \label{fig:pie3}
\end{figure}

\subsection{Answer to RQ1: Use Case Diagram Modeling} \label{sec:AnswerRQ1}

Based on the criteria defined in Table~\ref{tab:uccriteria} as well as Formula~(\ref{eq:eq1}) and Formula~(\ref{eq:eq2}), the UML models in the 45 project reports were evaluated and graded. The correctness rate for each criterion regarding use case modeling and its average score is shown in Table~\ref{tab:ucss}.


\begin{table}[h]
\caption{Scores of the created use case models}
\label{tab:ucss}
\centering
\begin{tabular}{p{6cm}p{1.5cm}}
\hline
\textbf{Criteria} &  \textbf{Correctness Rate (\%)}\\ \hline
\hline
UC1: Complete identification of actors & 31.11 \\ \hline
UC2: Complete identification of use cases & 88.89  \\ \hline
UC3: Accurate recognition of relationships among actors & 17.78 \\ \hline
UC4: Accurate recognition of relationships between actors and use cases & 100.00 \\ \hline
\textit{Average Score} & 59.44 \\ \hline
\end{tabular}%
\end{table}

As shown in Table~\ref{tab:ucss}, all types of modeling elements in the use case model can be identified with the help of LLM, including actors, use cases, the relationships between actors, and the relationships between actors and user cases, etc.
Particularly, the relationship between actors and use cases is the easiest to be correctly identified with the assistance of LLM, since the UML use case models in all the 45 project reports follow UC4 with a 100\% correctness rate. In addition, LLM performs excellently in identifying use cases, with a correctness rate of 88.89\%. Surprisingly, the complete identification of the relationships between actors seems to be the most difficult task in use case modeling for the LLM, with the lowest correctness rate of 17.78\%. Another challenge to LLM-aided use case modeling is to accurately identify all the actors in the given case study, as the correctness rate of UC1 is 31.11\%. For example, very few participants in our study successfully identified the generalization relationship between `Customer' and `GoldCustomer' in the given case study.


\subsection{Answer to RQ2: Class Diagram Modeling} \label{sec:AnswerRQ2}

The correctness rate for each criterion regarding class diagram modeling and its average score are shown in Table~\ref{tab:css}. It is observed that the LLMs gave much more help in the correct identification of classes, their operations, and the attributes, with correctness rates of 66.67\%, 75.56\%, and 91.11\% respectively. However, the correctness rate of accurately identifying relationships between classes is the lowest, accounting for 28.57\%.


\begin{table}[h]
\caption{Scores of the created class diagram}
\label{tab:css}
\centering
\begin{tabular}{p{6cm}p{1.5cm}}
\hline
\textbf{Criteria} &  \textbf{Correctness Rate (\%)}\\ \hline
\hline
CC1: Correct identification of classes & 66.67\\ \hline
CC2: Correct identification of operations & 75.56  \\ \hline
CC3: Correct identification of attributes & 91.11 \\ \hline
CC4: Accurate recognition of relationships among classes & 24.44 \\ \hline
\textit{Average Score} & 64.44 \\ \hline
\end{tabular}
\end{table}

\subsection{Answer to RQ3: Sequence Diagram Modeling} \label{sec:AnswerRQ3}
The correctness rate for each criterion with respect to sequence diagram modeling and its average score is shown in Table~\ref{tab:sss}.
We found that most of the three modeling elements in the sequence diagrams can be correctly identified with the help of LLM. Relatively, the correct identification of the messages between objects is lower than those of the other two modeling elements.


\begin{table}[h]
\caption{Scores of the created sequence diagram}
\label{tab:sss}
\centering
\begin{tabular}{p{6cm}p{1.5cm}}
\hline
\textbf{Criteria} &  \textbf{Correctness Rate (\%)}\\ \hline
\hline
SC1: Correct identification of objects & 73.33 \\ \hline
SC2: Correct identification of messages  & 68.89  \\ \hline
SC3: Correct identification of the order of messages & 82.22 \\ \hline
\textit{Average Score} & 74.81 \\ \hline
\end{tabular}
\end{table}

\subsection{Answer to RQ4: The Impact of Output Format on the Modeling Performance of LLMs} \label{sec:AnswerRQ4}

After reviewing the 45 project reports, the types of output formats requested by the participants during the human-LLM interaction can be categorized as follows:
\begin{itemize}
  \item \textbf{Auto-generated Diagram}: The output of the human-LLM interaction is directly used as the UML diagrams included in the project reports. That is, the participants are not involved in the representation of UML diagrams. Furthermore, there are two subtypes of these auto-generated diagrams.
  \begin{itemize}
      \item \textbf{Simple Wireframe}: The UML diagrams are composed of wires directly generated by LLM, as shown in Figure~\ref{fig:abc}(a). This output format may have a rudimentary appearance and a high probability of losing information.
      \item \textbf{PlantUML-based Diagram}: The UML diagrams in the project reports are automatically generated by PlantUML. In this way, the LLM is explicitly asked to give the PlantUML code of specified types of UML diagrams. This kind of output format can be the input of a specific tool, such as PlantUML, to automatically generate the graphical UML diagrams, as shown in Figure~\ref{fig:abc}(c).
  \end{itemize}
  \item \textbf{Hybrid-created Diagram}: This type of UML diagram in the project reports is the deduction of the raw output of the LLM. The raw output of the LLM can be either simple wireframes or textual descriptions. The participants can either manually improve or employ the prompt in LLM to optimize these raw outputs, and then draw the UML diagrams with specific UML modeling tools, such as StarUML. An example of hybrid-created diagram created with StarUML is shown in Figure~\ref{fig:abc}(b).
  
\end{itemize}

\begin{table}[h]
\caption{Scores of Three Types of Output Formats}
\label{tab:ofss}
\resizebox{\columnwidth}{!}{%
\begin{tabular}{ccccc}
\hline
\textbf{Output Format} & \textbf{Occurrence} & \multicolumn{2}{c}{\textbf{Score Distribution}} & \textbf{Average Score} \\ \hline
\multirow{3}{*}{Simple Wireframe}       & \multirow{3}{*}{4}  & [0, 6]   & 3 & \multirow{3}{*}{5.50} \\ \cline{3-4}
                                        &                     & [7, 9]   & 1 &                       \\ \cline{3-4}
                                        &                     & [10, 11] & 0 &                       \\ \hline
\multirow{3}{*}{PlantUML-based Diagram} & \multirow{3}{*}{16} & [0, 6]   & 3 & \multirow{3}{*}{6.94} \\ \cline{3-4}
                                        &                     & [7, 9]   & 13 &                       \\ \cline{3-4}
                                        &                     & [10, 11] & 0 &                       \\ \hline
\multirow{3}{*}{Hybrid-created Diagram} & \multirow{3}{*}{25} & [0, 6]   & 7 & \multirow{3}{*}{8.20} \\ \cline{3-4}
                                        &                     & [7, 9]   & 9 &                       \\ \cline{3-4}
                                        &                     & [10, 11] & 9 &                       \\ \hline
\end{tabular}
}
\end{table}

Table ~\ref{tab:ofss} shows the distribution of UML diagrams in the 45 project reports over these three types of output formats. Here, Formula~\ref{eq:eq3} defined in Section ~\ref{sec:rprocess} is applied to provide the total score of the UML diagrams, i.e., Score in the 3rd column of Table~\ref{tab:ofss}, in each project report. It was observed that hybrid-created diagram is the most used (25 out of the 45 reports) and higher quality (the average score is 8.26) output format in our work. Only two reports include simple wireframe diagrams, whose average score is the lowest, accounting for 5.5. The remaining 11 reports introduced PlantUML to automatically generate UML diagrams. Surprisingly, 81.25\% of these 16 participants (13 participants) got enough good UML models whose scores range from 7 to 9. The average score for these UML models in these 16 reports is 6.94. 

\section{Discussion} \label{sec:discussion}
In this section, we discuss the experimental results presented in Section~\ref{sec:results}. In Section ~\ref{sec:CI}, we perform a horizontal comparison of different sections within the experimental results to uncover more valuable information and draw further conclusions. We also interpret the results from the previous section, with possible reasons for their outcomes. In Section ~\ref{sec:implication}, we discuss how the results and conclusions from the aforementioned sections can be transformed to benefit software engineering educators, students, and relevant professionals. We also look ahead to the potential impact and significance that LLMs may have on the software engineering and requirements analysis industries in the future.

\subsection{Comparison and Interpretation}  \label{sec:CI}
Based on the experimental results and data from the previous section, we can conclude that LLMs can play a role in assisting with requirements analysis and modeling. However, shortcomings and instability also exist, and LLM's assistance does not guarantee that students with a basic understanding of requirements analysis can create fully compliant and correct UML models. This implies that at the current stage, LLMs cannot be considered a reliable tool for novice analysts in requirements analysis and software modeling.

\textbf{Results of RQ1, RQ2, and RQ3}: After comparing the results in Sections~\ref{sec:AnswerRQ1}, ~\ref{sec:AnswerRQ2}, and ~\ref{sec:AnswerRQ3}, we can identify a common trend: LLMs generally excel in recognizing specific objects (such as classes or use cases) from natural language text. However, extracting and analyzing relationships is not LLM's strong suit. Although it can still identify some elements related to types of relationships, their correctness and completeness are not guaranteed.

Furthermore, by comparing the average scores of the three types of UML models, we can see that LLM's modeling is more accurate for sequence diagrams. This may be because sequence diagrams involve fewer `relationship' elements and mostly consist of object-type elements that can be directly extracted and created. As a result, LLMs perform quite well in all three criteria for sequence diagrams.

\textbf{Results of RQ4}: The results in Section~\ref{sec:AnswerRQ4} highlight the excellent performance of hybrid-created diagrams among the three output formats. This further shows the advantage of human-drawn diagrams in software modeling when using LLMs. One possible reason is that LLMs are not trained to generate diagrams. Instead, its strength and primary function lie in natural language-based interactions. Therefore, content primarily in natural language should retain more information than outputs in the form of diagrams or code that can be transformed into diagrams. Furthermore, conveying information to humans first ensures that this information is reviewed and corrected before transformation, further amplifying the advantage of hybrid-created diagrams. However, we still observe that some hybrid-created diagrams have lower scores, possibly because of the influence and limitations of individual user expertise and experience.


\subsection{Implication} \label{sec:implication}
The answers to RQ1, RQ2, and RQ3 have demonstrated the feasibility of using large language model-based generative AI tools to assist in creating UML use case diagrams, class diagrams, and sequence diagrams. They have also exposed some of its shortcomings and limitations. The answers to RQ4 indicate possible ways to improve the accuracy and quality of modeling using LLMs under current conditions. Based on the research findings in this paper, there will be further exploration to address the challenges of using large language model tools for requirements acquisition and UML modeling, and more possibilities will be introduced into the field of requirements engineering and software engineering education.

\textbf{For educators}: The findings of this study inspire a new way to apply LLMs to the education and training of requirements analysis and modeling, providing new directions and goals for future software engineering education. Although at the current stage, there are still some challenges of using LLMs to generate UML models, it can give constructive and critical details and suggestions. This is undoubtedly promising to introduce LLMs in teaching activities. That is, using LLMs to evaluate and optimize models manually created by the students is also a promising direction. On the other hand, how to train the students to become software engineers who are proficient in using LLMs to assist in modeling work but not overly reliant on it will be a new proposition that software engineering educators need to discuss.

\textbf{For students}: The results demonstrated that LLMs can serve as an auxiliary tool to provide effective suggestions, appropriate guidance, and reference examples for software modeling. This can actually be helpful for beginners in software modeling. However, in the process of learning software modeling, students should also be aware of the weaknesses and limitations of LLMs and ensure that they have the ability to correct the occasional errors made by LLMs.

\textbf{For professionals}: This study highlights the obstacles that prevent large language models from becoming standard, powerful tools for requirements analysis and modeling that can be widely applied. These obstacles include weaknesses in identifying relationship-related elements and the issue of information loss when generating UML diagrams in certain formats. Requirements engineers who attempt to use LLMs to assist in their work need to be vigilant about the potentially serious impact of these issues. In the future, requirements engineers and AI researchers should work together to overcome these challenges, improve the performance of AI tools in software modeling, and create standardized tools that are truly suitable for this task.

\section{Limitations} \label{sec:limitations}
This exploratory study on the application of LLMs in UML modeling with novice analysts has some limitations.

\textbf{Experience of participants}: Since our analysis and conclusions are based on collected student reports, their individual abilities and experiences may have an impact on the final outcomes. Overall, most of the students who participated in this experiment were newcomers with relatively short training in requirements analysis and modeling, resulting in a generally lower skill level. Therefore, the conclusions drawn may not necessarily be applicable to knowledgeable and experienced industry professionals. On an individual level, despite these students completing the same course within the same timeframe, there are still differences in their grasp of requirements analysis and software modeling. Additionally, variations in students' attitudes toward course assignments may also influence our experimental results.

\textbf{Scoring of modeling results}: The experiment reports submitted by students were manually evaluated by team members, which may introduce subjectivity and bias, making it difficult to accurately assess the correctness of the modeling results. Additionally, in this study, for each criterion, we gave the score using a Boolean variable that was either 0 or 1. However, this scoring method may not accurately reflect the correctness level of each sample.

\textbf{Uncertainty of the LLM outputs}: Due to the inherent randomness exhibited by current generative AI based on LLMs during the interactions between users and LLMs, identical inputs may lead to different outputs, and its performance cannot be fully controlled with precision by the interactors. This may also have a negative impact on our experimental results and conclusions.


\section{Conclusions and Future Work} \label{sec:conclusion}

This paper explored how LLMs assist in creating three types of UML models: use case models, class diagrams, and sequence diagrams, as well as the potential factors affecting their performance. More specifically, we designed a modeling task and asked 45 undergraduate students who participated in a software requirements modeling course to create these three UML models. We required them to use LLM for assistance and submit experimental reports. By collecting and analyzing their reports, we found that LLM can play a role in assisting software modeling, but also has shortcomings and limitations. Additionally, we clarified the direction for optimizing LLM's recognition of modeling elements by comparing the impact of different output formats of UML diagrams.

As for the scoring problem mentioned in Section ~\ref{sec:limitations}, we plan to refine the scoring method for each criterion by using a numerically continuous variable as a score, thereby optimizing the accuracy of the scoring results.

We hope to expand the sample size to increase the credibility of our conclusions in the future. We also plan to extend the experiments to other types of UML models, such as activity diagrams and state diagrams. Additionally, we aim to conduct controlled experiments comparing the use of LLM for modeling assistance with cases where LLM is not used at all. This quantitative exploration will help us understand the extent to which LLM can assist in improving the quality of UML models. Furthermore, considering the rapid development of large language models in China, such as ERNIE Bot and Qwen, these models will also be included in the scope of future research.

\section*{Acknowledgments} \label{sec:ack}
This research work has been supported by the National Natural Science Foundation of China (NSFC) with No. 62172311.

\balance
\bibliographystyle{ieeetr}
\bibliography{main}

\end{document}